\documentclass[12pt,preprint,flushrt]{aastex}

\newcommand{\simgt}{\lower.5ex\hbox{$\; \buildrel > \over \sim \;$}}
\newcommand{\simlt}{\lower.5ex\hbox{$\; \buildrel < \over \sim \;$}}
\newcommand{\bm}[1]{\mbox{{\it \boldmath$#1$}}}

\newcommand{\skaco}[1]{\langle{#1}\rangle}
\newcommand{\colskip}{@{\hspace{0.3in}}}
\newcommand{\baredth}{\;\overline{\raise1.0pt\hbox{$'$}\hskip-6pt
\partial}\;}
\newcommand{\edth}{\;\raise1.0pt\hbox{$'$}\hskip-6pt\partial\;}
\newcommand\etal{{\it et al.}}

\newcommand{\gsim}{\lower.5ex\hbox{$\; \buildrel > \over \sim \;$}}
\newcommand{\lsim}{\lower.5ex\hbox{$\; \buildrel < \over \sim \;$}}
\begin{document}

\title{Color Tomography}

\author{Bhuvnesh Jain$^1$, Andrew Connolly$^2$ and Masahiro Takada$^3$}
\affil{
{}$^1$Dept. of Physics and Astronomy, University of Pennsylvania,
Philadelphia, PA 19104\\
{}$^2$Dept. of Physics and Astronomy, University of Pittsburgh, 
Pittsburgh, PA 15260\\
{}$^3$Dept. of Physics and Astronomy, Tohoku University, Sendai
980-8578, Japan}
\email{
bjain@physics.upenn.edu,ajc@phyast.pitt.edu,takada@astr.tohoku.ac.jp}

\begin{abstract}
Lensing tomography with multi-color imaging surveys can probe dark
energy and the cosmological power spectrum. However accurate photometric
redshifts for tomography out to high redshift 
require imaging in five or more bands, which is
expensive to carry out over thousands of square degrees.  Since
lensing makes coarse, statistical use of redshift information, we explore the 
prospects for tomography using limited color information from 
two or three band imaging. With an appropriate calibration sample, 
we find that it is feasible to create up to four redshift 
bins using imaging data in just the $g$, $r$ and $i$ bands. 
We construct such redshift sub-samples from mock catalogs 
by clustering galaxies in color space and
discarding regions with poorly-defined redshift distributions. The 
loss of galaxy number density decreases the accuracy of lensing 
measurements, but even losing half or more of the galaxies 
is not a severe loss for large area surveys. We estimate the errors 
on lensing power spectra and dark energy parameters with color 
tomography and discuss trade-offs in survey area and filter choice. 
We discuss the systematic errors that may change our conclusions,
especially the information needed to tackle intrinsic alignments. 
\end{abstract}

\keywords{cosmology:gravitational lensing --- cosmology:observation}

\section{Introduction}

Over the last decade a concordance model has emerged in cosmology
in which about two-thirds of the energy density of the universe today 
may be in the form of ``dark energy''. This explains
the observation that we reside in an accelerating universe
\citep{Riess,Perl}.  Despite its importance to the formation and
evolution of the universe there are no compelling theories that
explain the energy density nor the properties of the dark energy. 

To address questions about the nature of dark energy a number of
ambitious wide-field optical and infrared imaging surveys have been
proposed. These range from space-based missions in the optical and
infrared, such as the 
Supernova Acceleration Probe 
(SNAP\footnote{http://www.snap.lbl.gov}, proposed 
as the space-based Joint Dark Energy Mission (JDEM)), to ground-based
surveys such as the Panoramic Survey Telescope \& Rapid Response
		System 
(Pan-STARRS\footnote{http://pan-starrs.ifa.hawaii.edu}),
the Dark Energy Survey (DES\footnote{
http://www.darkenergysurvey.org}), the Large Synoptic Sky Survey
(LSST\footnote{http://www.lsst.org}) and others. Each of these
missions approaches the study of dark energy using multiple,
complementary observational probes: gravitational weak lensing (WL) to
study the growth of structure and geometry, baryon oscillations to
measure the angular diameter distance vs. redshift relation and Type Ia
supernovae to measure the  luminosity distance vs. redshift relation.

In this paper we focus on one of these probes, weak lensing, and how
the design of a survey might impact its scientific value for
constraining dark energy. Observations to date have succeeded in
measuring the amplitude $\sigma_8$ of the $z\approx0$ dark matter
power spectrum to $\approx10\%$ accuracy ({\it cf.}
\citet{Jarvis05,CFHLS05} and references therein) by surveying $\simeq
100$~deg$^2$ of sky. To constrain the equation of state of dark energy
requires that we improve the accuracy of these measures by close to an
order of magnitude. 
To achieve this deep imaging surveys have been proposed that
cover surveys of 1,000-20,000 square degrees in 5-6 filters. There is,
however, a natural trade off in the design of these surveys; depth vs
number of filters vs area surveyed. To date the appropriate
weighting of these components is not fully understood. For example,
given the time and cost of a survey, does a large area, shallow survey
in a small number of filters provide more scientific return than a
deep survey in a large number of filters but sampling only a few
hundred square degrees?

One of the primary requirements for any lensing application is the
need for redshift information on the source galaxies. In the first
studies of lensing only the statistical redshift distribution was
available, but now surveys aim to get photometric redshifts (Connolly
\etal \ 1995; hereafter photo-z's) for individual galaxies. Accurate
photo-z information enables new qualitative and quantitative advances
in lensing. Conversely photo-z errors can be a limiting systematic in
the use of lensing for precision cosmology.

Lensing tomography refers to the use of depth information in the
source galaxies to get three-dimensional information on the lensing
mass \citep{Hu99}. By binning source galaxies in photo-z bins, the
evolution of the lensing power spectrum can be measured. 
This greatly improves the sensitivity of lensing to
dark energy in cosmological applications. 
The lensing power spectra measured from two redshift bins is shown in
Figure \ref{fig:cl}. The two auto-spectra and the cross-spectrum,
along with expected statistical errors, are shown.  The relative shift in
the amplitudes of the lensing spectra is sensitive to the 
properties of dark energy.  It depends on both distances
and the growth of structure, thus enabling tests of dark energy or
modified gravity explanations for the cosmic acceleration. In Figure
\ref{fig:cl}, the amplitude shifts of the three spectra (for a
given cosmological model) depend on the mean redshifts and widths of the two
photo-z bins; clearly any errors in the bin redshifts will degrade the
ability to discriminate cosmological models.

Thus the use of broadband
multicolor photometry to estimate the distances of galaxies (i.e.\
photometric redshifts; henceforth abbreviated as photo-z's) is 
critical to the use of lensing for dark energy studies. 
The capability of these surveys to meet their
scientific goals will depend on our ability to characterize
the uncertainties present within photometric redshift estimates,
i.e. the scatter, bias and fraction of outliers. 
Photometric redshifts must be calibrated with an appropriate sample of 
spectroscopic redshifts (Huterer \etal \ 2005; Ma \etal \ 2005). 
This may be done more cheaply by using 
auto- and cross-correlations of photometric and
spectroscopic redshifts samples \citep{Newman2006,Schneider2006}, which
can also be used to estimate the redshift distribution 
for a galaxy sample where the calibration data is incomplete 
(see also Zhan \& Knox 2006). 

In this paper we study the possibility of doing lensing tomography from 
limited color information. Imaging in five or more bands is needed for 
well-characterized photo-z's of galaxies extending beyond $z\sim 1$. We 
consider here whether lensing tomography can be carried out 
from a wide-area imaging survey in just two or three filters, 
along with a relatively small calibration sample 
that provides the statistical redshift distributions in all parts of
the color space. These can be used to create a few sub-samples of 
the full galaxy sample that occupy distinct redshift bins, 
while discarding galaxies with colors that lead
to badly defined redshift distributions. 
Since lensing does not require redshift bins much 
narrower than $0.2$-$0.4$ in redshift, and does not need a full or fair
sample of the galaxy population, there is reason to expect that
limited color information may be sufficient for tomography. 
We use a mock catalog of galaxies with known redshifts to see
how well one can make cosmological measurements with this approach, 
taking into account statistical errors in lensing measurements. 
In a future paper we will expand upon this analysis to 
include various sources of systematic errors in colors/photo-z's and
in the lensing measurements.  

In Section 2 we describe the formalism for computing lensing power spectra
given the redshift distribution of a galaxy sample. In Section 3 the mock 
catalog used for our study is described. Section 4 contains the results on 
how well one can do with color cuts in making well separated redshift 
distributions for galaxy sub-samples. The errors on the power spectra and
dark energy parameters are compared with the forecasts for idealized
photo-z's. In Section 5 we discuss the prospects for tomography with
color cuts from imaging in a limited number of filters -- which we call 
color tomography. 

\section{Lensing Formalism}

We will use the shear power spectrum as the lensing statistic for dark
energy constraints. Similar results can be obtained using two-point
correlations and the mass aperture statistic. The key element in
constraining dark energy is to
use source galaxies in different redshift bins to probe the
evolution of mass fluctuations and the geometric factors involved in
lensing. 

\subsection{Preliminaries: cosmology and weak lensing}
\label{cosmo}

We work in the context of spatially flat cold dark matter models for
structure formation.  The expansion history of the universe is given by
the scale factor $a(t)$ in a homogeneous and isotropic universe.  
The expansion rate, the Hubble parameter, is
given in terms of the matter density $\Omega_{\rm m}$ (the
cold dark matter plus the baryons) and dark energy density $\Omega_{\rm
de}$ at present (in units of the critical density $3H_0^2/(8\pi G)$, 
where $H_0=100~ h~{\rm km}~ {\rm s}^{-1}~ {\rm Mpc}^{-1}$ is the Hubble
parameter at present) by
\begin{equation}
H^2(a)=H_0^2\left[\Omega_{\rm m}a^{-3}+\Omega_{\rm de}
e^{-3\int^a_1 da' (1+w(a'))/a'},
\right]
\end{equation}
where we have employed the normalization $a(t_0)=1$ today
and $w(a)$ specifies the equation of state for dark energy as
\begin{equation}
w(a)\equiv \frac{p_{\rm de}}{\rho_{\rm de}}=-\frac{1}{3}
\frac{d\ln \rho_{\rm de}}{d\ln a}-1. 
\end{equation}
Note that $w=-1$ corresponds to a cosmological constant.  The comoving
distance $\chi(a)$ from an observer at $a=1$ to a source at $a$ is
expressed in terms of the Hubble parameter as
\begin{equation}
\chi(a)=\int^1_a\!\!\frac{da'}{H(a')a^{\prime 2}}. 
\end{equation}
This gives the distance-redshift relation $\chi(z)$ via the relation
$1+z=1/a$.

Next we need the growth of density perturbations. 
In linear theory, all Fourier modes of the mass density perturbation, 
$\delta(\equiv \delta
\rho_m/\bar{\rho}_m)$, grow at the same rate, the growth rate $D(a)$:
$\tilde{\delta}_{\bm{k}}(a)\propto D(a)$. 
Note that we use the primordial curvature power spectrum amplitude to
normalize the linear 3D mass power spectrum, and therefore
$D(a)$ is normalized as $D(a_{\rm md})/a_{\rm md}=1$
in the deeply matter dominated regime (e.g., $a_{\rm md}=10^{-3}$; 
see equation~(10) in Takada 2006 for details). 
In the following, the tilde symbol is used to denote Fourier components. 

The shear power spectrum is identical to that of the convergence,
which is easier to work with as it is a scalar. 
In the context of cosmological gravitational lensing, the convergence
field is expressed as a weighted projection of the three-dimensional
density fluctuation field between source and observer (e.g., see 
Bartelmann \& Schneider 2001; Mellier 1999 for reviews):
\begin{equation}
\kappa(\bm{\theta})=\int_0^{\chi_H}\!\!d\chi W(\chi) 
\delta[\chi, \chi\bm{\theta}],
\label{eqn:kappa}
\end{equation}
where $\bm{\theta}$ is the angular position on the sky, $\chi$ is the
comoving distance, and $\chi_H$ is the distance to the horizon.  Note
that for a flat universe the comoving angular diameter distance is
equivalent to the comoving distance.  
The lensing weight function $W(\chi)$ is defined in equation
\ref{eqn:weight} below.
%%
%\begin{eqnarray}
%W(\chi)=\frac{3}{2}\Omega_{\rm m0}H_0^2 a^{-1}(\chi)
%\chi
%\frac{1}{\bar{n}_{\rm g}}\int^{\chi_H}_\chi\!\!d\chi_s~ 
%p_s(z)\frac{dz}{d\chi_s} 
%\frac{\chi_{\rm s}-\chi}{\chi_s},
%\label{eqn:weightgl}
%\end{eqnarray}
%
%%
%\begin{equation}
%\bar{n}_{\rm g}=\int_0^{\chi_H}\!\!d\chi_s~ p_s(z)\frac{dz}{d\chi_s}. 
%\end{equation}
%%
%%%%%%%%%%%%%%%%%%%%%%%%%%%%%%%%%%%%%%%%%%%%%%%%%%%%%%%%%%%%%%%%%%%%%%

Photometric redshift information on source
galaxies allows us to subdivide the galaxies
into redshift bins. The average number density of galaxies in a
redshift bin $i$, defined to lie between the comoving distances 
$\chi_i$ and $\chi_{i+1}$, is given by
\begin{equation}
\bar{n}_i=\int_{\chi_i}^{\chi_{i+1}}\!\!d\chi_s~ p_s(z)\frac{dz}{d\chi_s}. 
\label{eqn:ni}
\end{equation}
where $p_s(z)$ is the redshift selection function of source
galaxies. It is normalized as
$\int_0^{\infty}\!\!dz ~ p(z)= \bar{n}_{\rm g}$, where 
$\bar{n}_{\rm g}$ is the average number density per unit steradian.
Note that $\bar{n}_i$ determines the shot noise contamination
due to the intrinsic ellipticities of galaxies 
for the power spectrum measurement in the
$i$ bin (see equation (\ref{eqn:covps}) and discussion below). 
The convergence field for subsample $i$ is given by using in equation
\ref{eqn:kappa} 
%%
%\begin{equation}
%\kappa_{(i)}\!(\bm{\theta})=\int_0^{\chi_H}\!\!d\chi~ W_{(i)}(\chi) 
%\delta[\chi, \chi\bm{\theta}],
%\label{eqn:kappai}
%\end{equation}
%
the lensing weight function $W_{(i)}$, given by
\begin{eqnarray}
W_{(i)}(\chi)&=&
\left\{
\begin{array}{ll}
{\displaystyle 
\frac{W_0}{\bar{n}_i}
a^{-1}\! (\chi)~ \chi
\int_{\chi_{i}}^{\chi_{i+1}}\!\!d\chi_{s}~ 
p_s(z)\frac{dz}{d\chi_s} \frac{\chi_{\rm s}-\chi}{\chi_s}},
& \chi\le\chi_{i+1},\\
0,& \chi>\chi_{i+1}.
\end{array}
\right.
\label{eqn:weight}
\end{eqnarray}
where $W_0=3/2\ \Omega_{\rm m0}H_0^2$. 
%We have ignored possible errors in
%the photometric redshifts for simplicity.
How a dynamically evolving dark energy model changes the lensing weight
function is shown in Figure 3 in Huterer (2002). 
For example, increasing $w$ lowers $W_{(i)}$ --- 
similar to the dependence of the growth rate of mass clustering 
for CMB normalization. 
Thus the dependence of lensing observables
on the equation of state arises roughly equally from the two effects. 

\subsection{The lensing power spectrum and its covariance}

To compute the convergence power spectrum, we employ
the flat-sky, Limber equation
which is a good approximation over angular scales of 
interest (Kaiser 1992; Hu 2000). Within this framework the lensing convergence 
field is decomposed into angular modes based on the two-dimensional Fourier
transform: $\kappa(\bm{\theta})=\sum_{\bm{l}}
\tilde{\kappa}_{\bm{l}}e^{i\bm{l}\cdot\bm{\theta}}$. The angular 
power spectrum, $C(l)$, is defined as
\begin{eqnarray}
\skaco{\tilde{\kappa}_{\bm{l}_1}\tilde{\kappa}_{\bm{l}_2}}
&=&(2\pi)^2\delta^D(\bm{l}_{12})C(l_1),
\end{eqnarray}
where $\delta^D(\bm{l})$ is the Dirac delta function, $\skaco{\cdots}$
denotes ensemble averaging, and
$\bm{l}_{12}=\bm{l}_1+\bm{l}_2$.

\begin{figure}[t]
\epsscale{0.6}  
\plotone{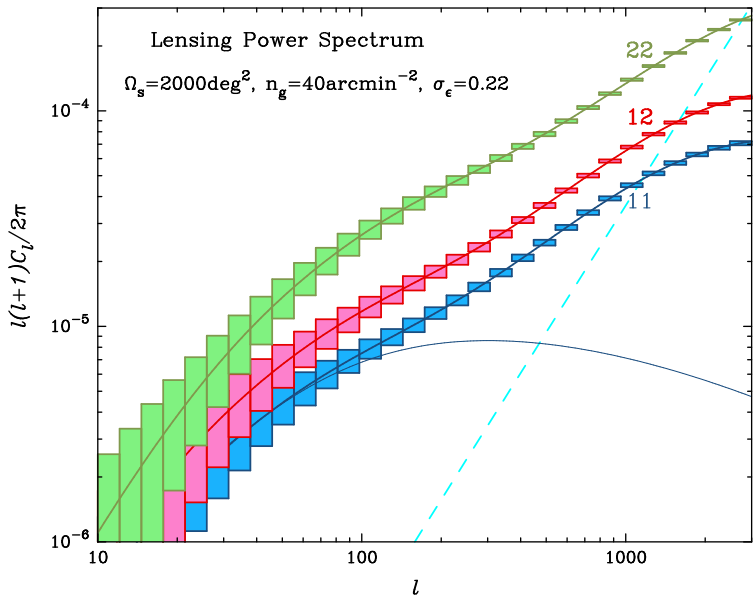}
\epsscale{1.0}  
\caption{The lensing auto- and cross-power spectra for galaxies in two
 redshift bins, $0\le z_1\le 1.3$ and $1.3\le z_2$. 
The solid curves are the results for the $\Lambda$CDM model, 
computed from the Smith03 fitting formula. The boxes show the expected 
measurement error due to sample variance
and intrinsic ellipticities. The linear 
auto-power spectrum for the low-$z$ bin 
is shown by the thin solid line to show how significant
the non-linear effect is. The thin dashed curve shows the shot noise 
contribution to the power.  }
 \label{fig:cl}
\end{figure}

For lensing tomography, we use all the auto- and cross-power spectra
that are constructed from source galaxies divided into redshift bins.
The angular power spectrum between redshift bins $i$ and $j$, $C_{(ij)}(l)$,
is given by
\begin{eqnarray}
C_{(ij)}(l)
=\int^{\chi_H}_0\!\!d\chi
W_{(i)}\!(\chi) W_{(j)}\!(\chi)
\chi^{-2}~ P_\delta\!\left(k=\frac{l}{\chi}; 
\chi\right),
\label{eqn:cltomo}
\end{eqnarray}
where the lensing weight function $W_{(i)}$ 
is given by equation (\ref{eqn:weight}) 
and $P_\delta(k)$ is the three-dimensional mass power spectrum. 
Using $n_s$ redshift bins
leads to $n_s(n_s+1)/2$ cross and auto power spectra. 
%Note that we have used Limber's equation (e.g. Kaiser 1992), 
%which is a good approximation over the angular modes we consider, 
%with $50\le l\le 3000$, corresponding to angular scales between $5'$ and a
%few degrees.  
The non-linear gravitational evolution
of $P_\delta(k)$ significantly enhances the amplitude of the lensing
power spectrum on angular scales $\ell\simgt 100$ 
(see Figure \ref{fig:cl}).  Therefore, we need an accurate model of 
$P_\delta(k)$, for which we employ the fitting formula proposed
by Smith et al. (2003, hereafter Smith03).  We assume that
the Smith03 formula can be applied to dark energy cosmologies, if we
replace the growth factor in the formula with that for a given dark
energy cosmology (White \& Vale 2004; Linder \& White 2005). 
%%linder and white??
The issue of how accurately the non-linear power spectrum needs to 
be calibrated
to attain the full potential of lensing surveys is addressed in Huterer
\& Takada (2005). 

%%%%%%%%%%%%%%%%%%%%%%%%%%%%%%%%%%%%%%%%%%%%%%%%%%%%%%%%%%%%%%%%%%%%%%
%%%%%%%%%%%%%%%%%%%%%%%%%%%%%%%%%%%%%%%%%%%%%%%%%%%%%%%%%%%%%%%%%%%%%%
Measured shear correlations contain a shot-noise contribution from the 
intrinsic ellipticities of source galaxies. Assuming that the ellipticity 
distribution is uncorrelated between different galaxies, the observed 
power spectrum between redshift bins $i$ and $j$ can be expressed as 
(Kaiser 1992, 1998; Hu 1999) 
\begin{eqnarray}
C^{\rm obs}_{(ij)}(l)
=C_{(ij)}(l)+\delta_{ij}\frac{\sigma_{\epsilon}^2}{\bar{n}_{i}},
\label{eqn:obscl}
\end{eqnarray}
where $\bar{n}_i$ is the average number density of galaxies in redshift
bin $i$, as given by equation (\ref{eqn:ni}), and $\sigma_{\epsilon}$ is
the intrinsic shape noise of each galaxy. 
The Kronecker delta
function enforces the fact that the cross power spectrum with $i\ne
j$ is not affected by shot noise (Hu 1999).  In this sense, the
cross-power spectrum is an unbiased estimator of the cosmological
signal.  We have ignored other possible contaminations such as 
observational systematics and intrinsic ellipticity alignments.
% (the
%latter is also likely to be negligible for cross power spectra). 

The power spectrum covariance is needed to understand
statistical errors on the power spectrum measurement. 
The covariance between the power spectra $C_{(ij)}(\ell)$ and
$C_{(mn)}(\ell')$ is 
\begin{eqnarray}
{\rm Cov}[C^{\rm obs}_{(ij)}(l),C^{\rm obs}_{(mn)}(l')]
&=&\frac{2\delta_{ll'}}{(2l+1)\Delta l f_{\rm sky}} 
\left[C^{\rm obs}_{(im)}(l)C_{(jn)}^{\rm obs}(l)
+C^{\rm obs}_{(in)}(l)C_{(jm)}^{\rm obs}(l)\right]
\label{eqn:covps}
\end{eqnarray}
where $f_{\rm sky}$ is the fraction of sky 
covered and $\Delta \ell$ is the bin width centered at $\ell$. 
%, so that
%the area of the shell is $A(l)=2\pi \ell \Delta \ell$. 
We have used only
the Gaussian contribution to the covariance which does not
lead to any correlation between the power spectra of different $\ell$ 
modes. We restrict our analysis to angular scales $\ell\le
3000$; as discussed below, our conclusions are stronger 
with the inclusion of non-Gaussian covariances or 
if a more conservative 
(lower) choice of the maximum $\ell$ is employed.   

Figure \ref{fig:cl} shows the lensing power spectra for two redshift
bins, leading to 3 different power spectra as indicated. The
solid curves are the results from the Smith03 fitting formula. To 
estimate the errors on the measured power spectra, we
parameterized a lensing survey by its survey area, $2,000$ degree$^2$,
%sky coverage 
%$f_{\rm sky}=0.1$ ($\approx 4000$ degree$^2$), 
the galaxy number density $\bar{n}_{\rm g}=40~ {\rm
arcmin}^{-2}$ and the rms of intrinsic ellipticities
$\sigma_{\epsilon}=0.22$. The number density we have used would be achievable 
with a limiting magnitude $r\gsim 25$ imaging survey in excellent seeing
conditions. The dashed line in Figure
\ref{fig:cl} shows the contribution from intrinsic ellipticity shot
noise to the power spectrum errors. 
%%% MT
For the power spectra shown, the
shot noise contribution becomes smaller than the sample  variance
at wavenumbers $\ell$ 
smaller than the intersection of the power spectrum points 
with the shot noise line. 
%%%%%%%%%%%%
%It is clear that the
%power spectrum for higher redshift bin has greater amplitude because of
%the greater lensing efficiency described by the lensing weight function 
%$W_{(i)}$ (see equation (\ref{eqn:weight})).  
The correlation coefficient
between the power spectra of the redshift bins,
$R_{ij}=C_{(ij)}(l)/[C_{(ii)}(l)C_{(jj)}(l)]^{1/2}$, quantifies how the
power spectra are correlated. Even with only two
redshift bins, the power spectra are highly correlated ($R_{12}\sim
0.8$). One thus gains little information from fine subdivisions of the
redshift bins (Hu 1999, 2002a,b).  The box around each curve shows the
expected measurement error at a given bin of $\ell$, 
which includes  sample variance and the error due to 
intrinsic ellipticities. The sample variance dominates
the error over much of the angular scales that provide 
cosmological information.  
Finally, to clarify the effect of non-linear gravitational clustering, 
the thin solid curve shows the prediction of $C_{(11)}(l)$ from the
linear mass power spectrum: non-linear evolution significantly
enhances the amplitude for $\ell\simgt 100$ (Jain \& Seljak 1997).

\begin{figure}[t]
\epsscale{0.6}  
\centering
\plotone{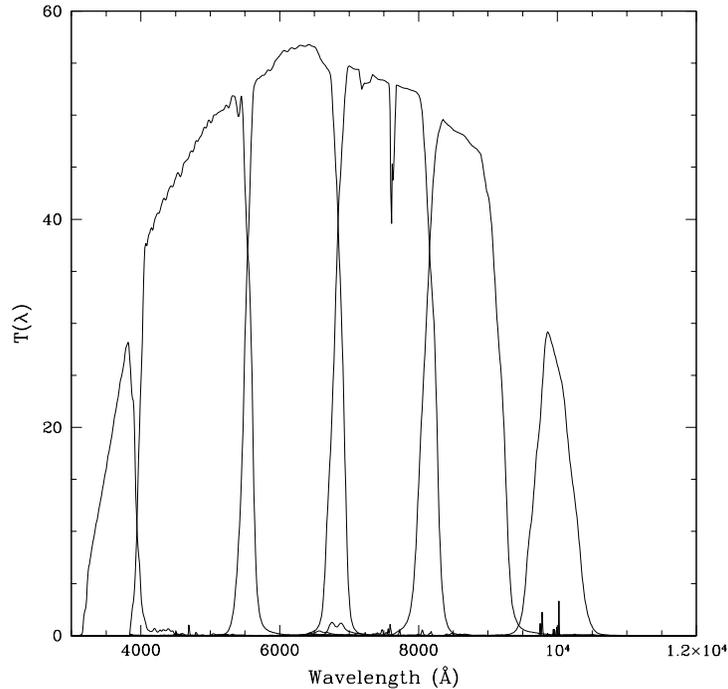}
\epsscale{1.0}  
\caption{\small The filter response functions in the
$u,g,r,i,z$ and $y$ filters. Note the high transmission 
efficiency in the $g,r,i$ filters compared to $u$ and $y$. }
\label{fig:filters}
\end{figure}

\begin{figure}[t]
\centering
\plottwo{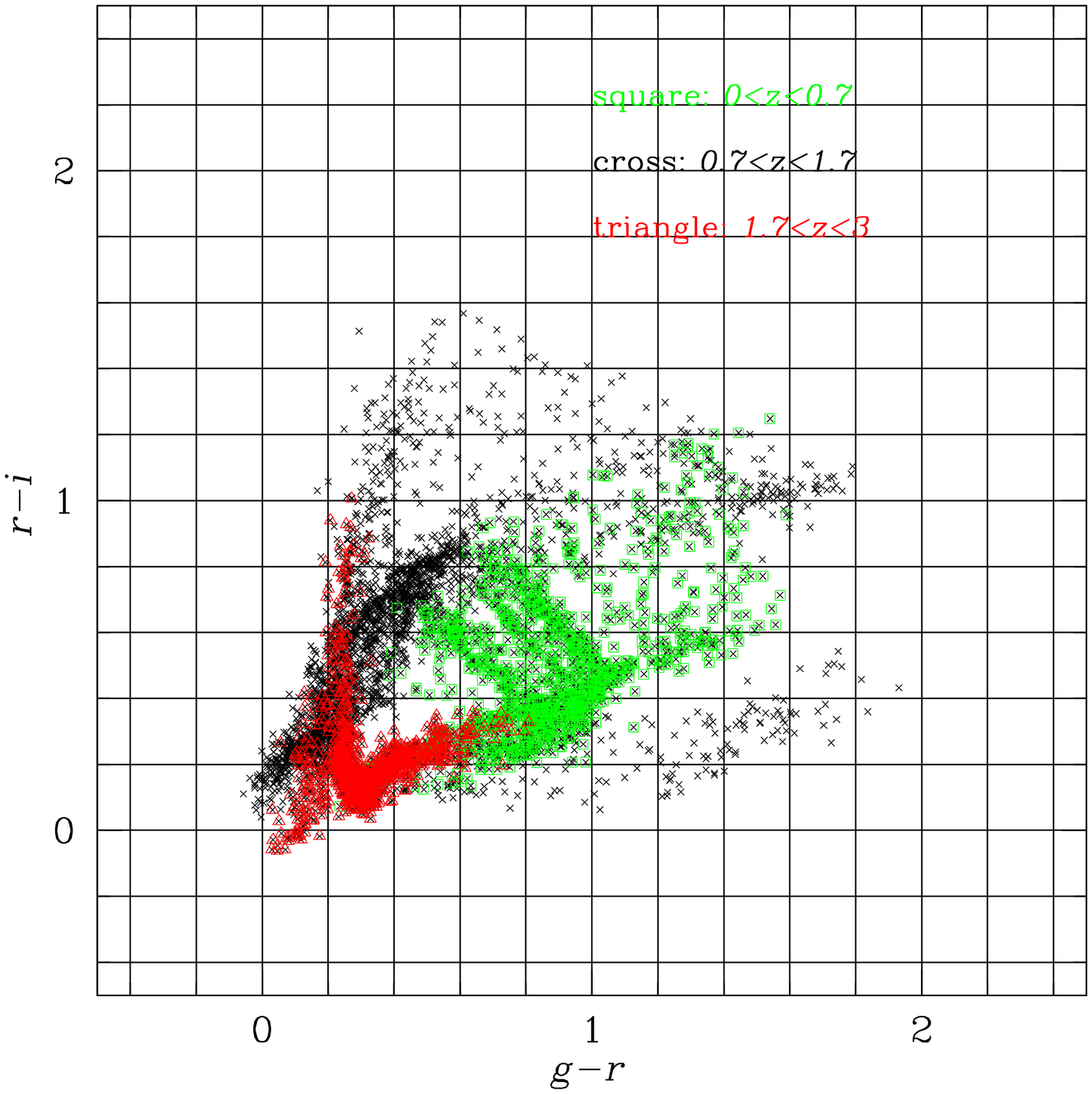}{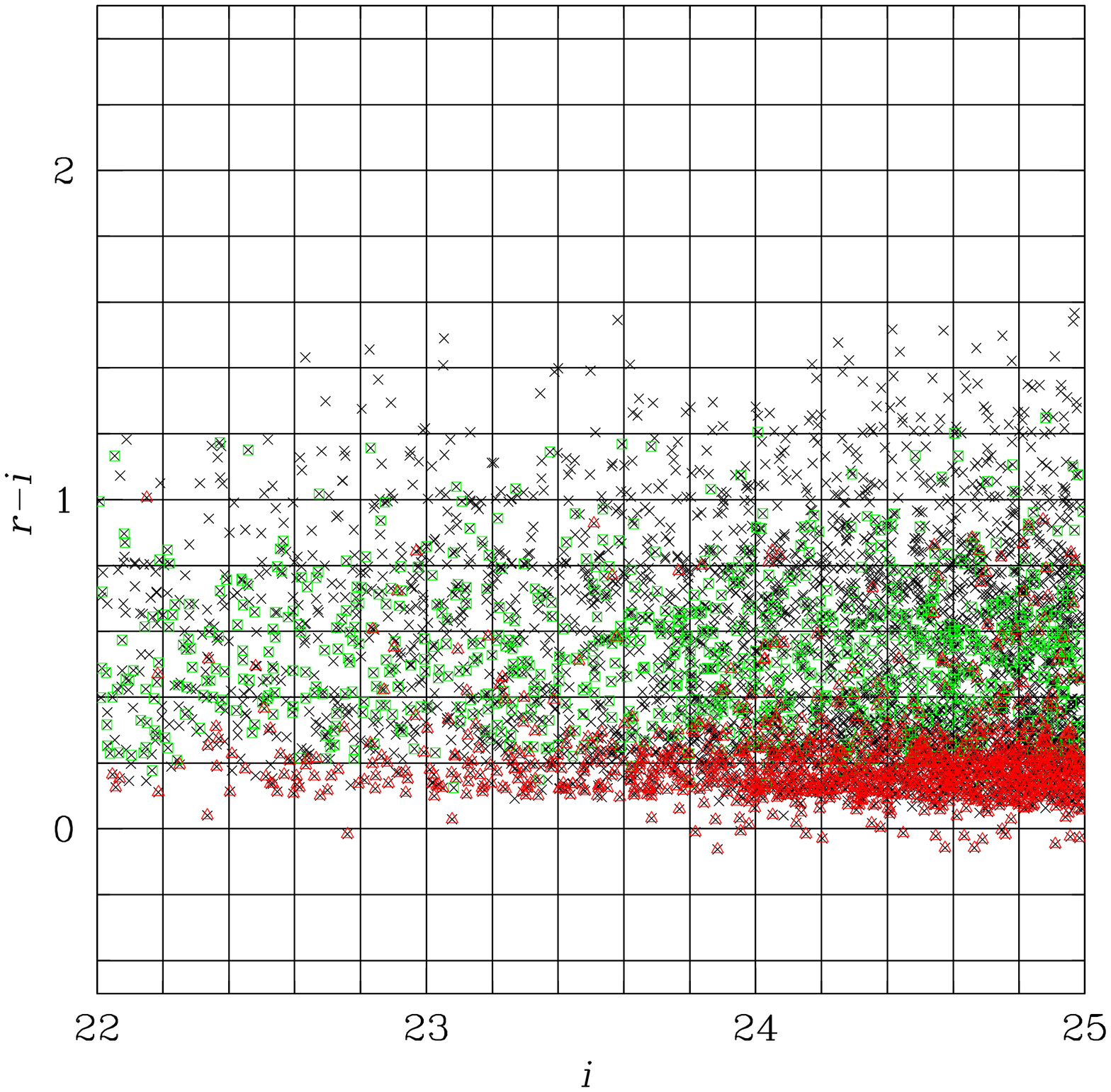}
\caption{\small The left panel shows the distribution of galaxy colors
($r-i$ vs. $g-r$)  in a simulated catalog with limiting magnitude $r=25$. 
The right panel shows $r-i$ color vs. $i$ magnitude. 
It is clear that the regions in color space occupied by galaxies
at different redshifts are not distinct, and cannot be separated by
orthogonal color cuts. However by selecting arbitrarily shaped regions 
in color space with well defined redshift distributions, we can attempt to
create distinct redshift distributions as shown in the following
figures. 
}
\label{fig:colors}
\end{figure}

\section{Mock galaxy catalogs}

A Monte-Carlo realization of $10^6$ galaxies was generated to
approximate the redshift and colors distributions of galaxy samples
obtained by the next generation of photometric surveys (e.g. the
LSST). Galaxies were initially selected to match the observed
number-magnitude relation for $i$-band selected galaxies of Metcalfe
{\it et al.}  (2001) to a depth of $i$=27. For each galaxy a redshift
and spectral type was assigned. The redshifts were drawn from the
magnitude-redshift distributions observed by the DEEP2 spectroscopic
survey (Willmer {\it et al.} 2006) and the evolution of the distribution
of spectral types of galaxies was based on the observations and models
of Franceschini {\it et al.} 2006.

%The spectral types were assigned based on the redshift 
%dependent distributions given in (ref??). 

Colors were estimated, as a function of redshift and spectral type,
using the $u,g,r,i,z,y$ filter response functions of the LSST (see
Figure \ref{fig:filters}). In total 10 spectral templates were used in
this sample which were derived from the observed spectral properties of
galaxies in the SDSS (including emission lines) and supplemented with a
young (50 Myr) star forming galaxy template drawn from the models of
Bruzual and Charlot (2003). Photometric uncertainties are estimated
based on the LSST exposure calculator assuming that each source has been
observed approximately 400 times in each filter. The final catalog was
limited to $r<25$ at which depth the galaxies have a signal-to-noise of
approximately 15 for LSST's exposure times. Thus the photometric errors
in our catalog are smaller than would be expected in shallower surveys;
we leave a detailed modeling of this and other errors for future work.

Figure \ref{fig:colors} shows the distribution of galaxies in color
space. Galaxies in three redshift intervals are shown to illustrate
that, while they generally occupy different regions in color space,
there is significant overlap, and moreover they cannot be isolated by
making simple color cuts. We use these properties to guide our algorithm
for selecting regions in color space for tomography in the next section.

\begin{figure}[t]
\plottwo{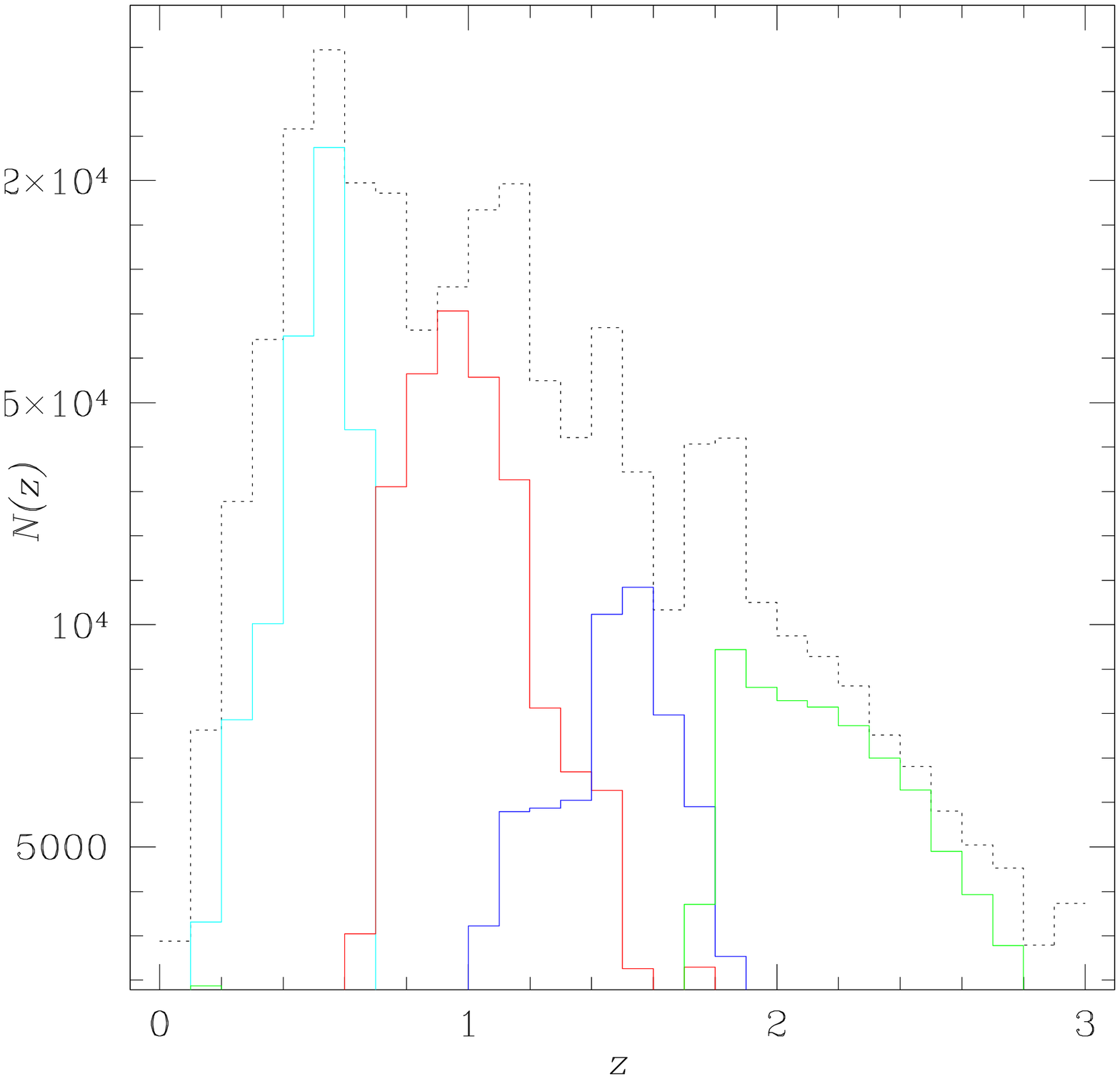}{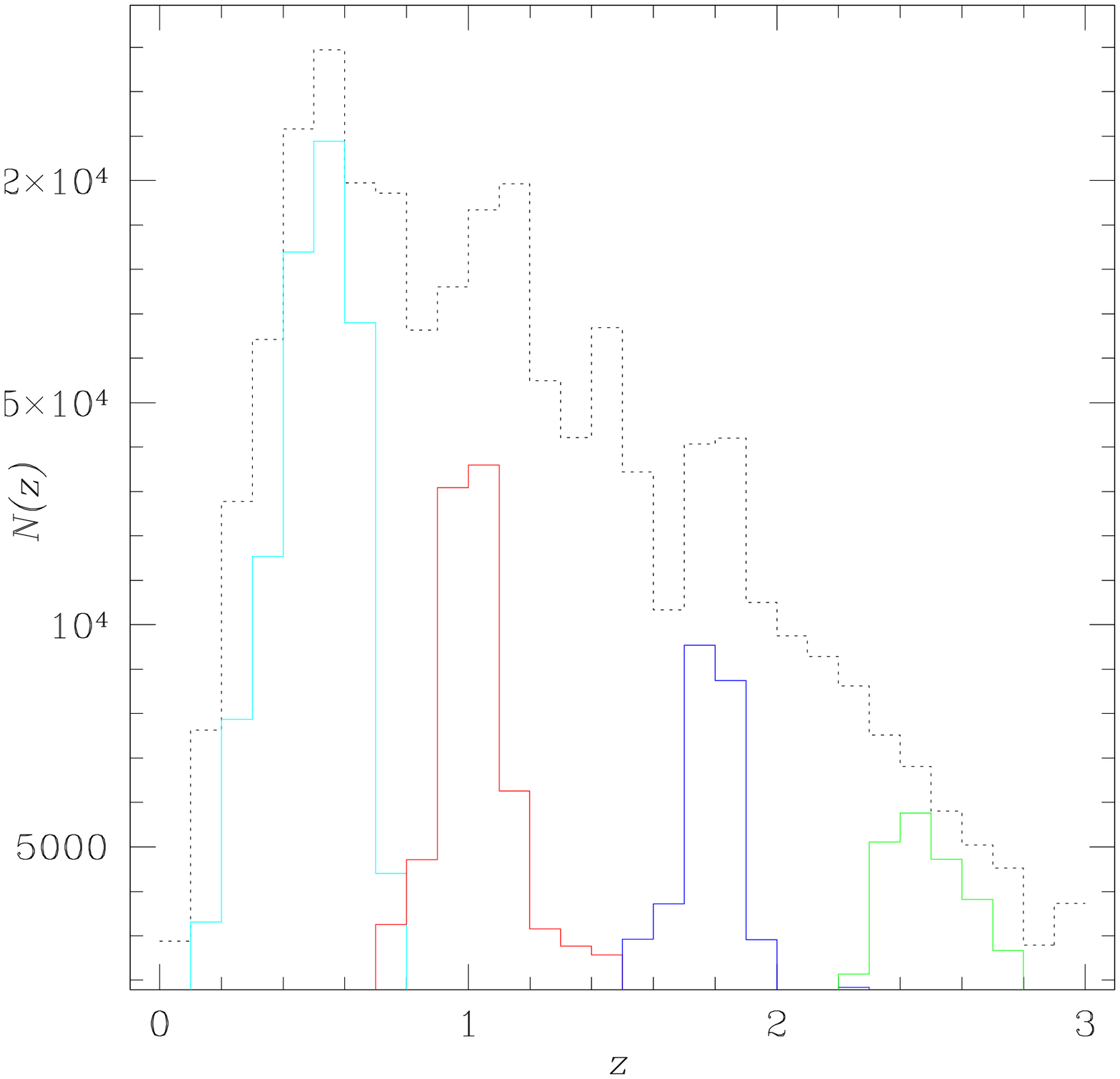}
\caption{\small The left panel shows the redshift distribution of 
four galaxy sub-samples using cuts in $r-i$ and $g-r$. The full sample
is shown by the black lines. It is clear 
that the three sub-samples have overlap, but only at the $\sim 10\%$ level. 
The right panel shows the distributions obtained with more drastic
cuts to better isolate the redshift sub-samples. 
}
\label{fig:dndz}
\end{figure}

\begin{figure}[t]
\plottwo{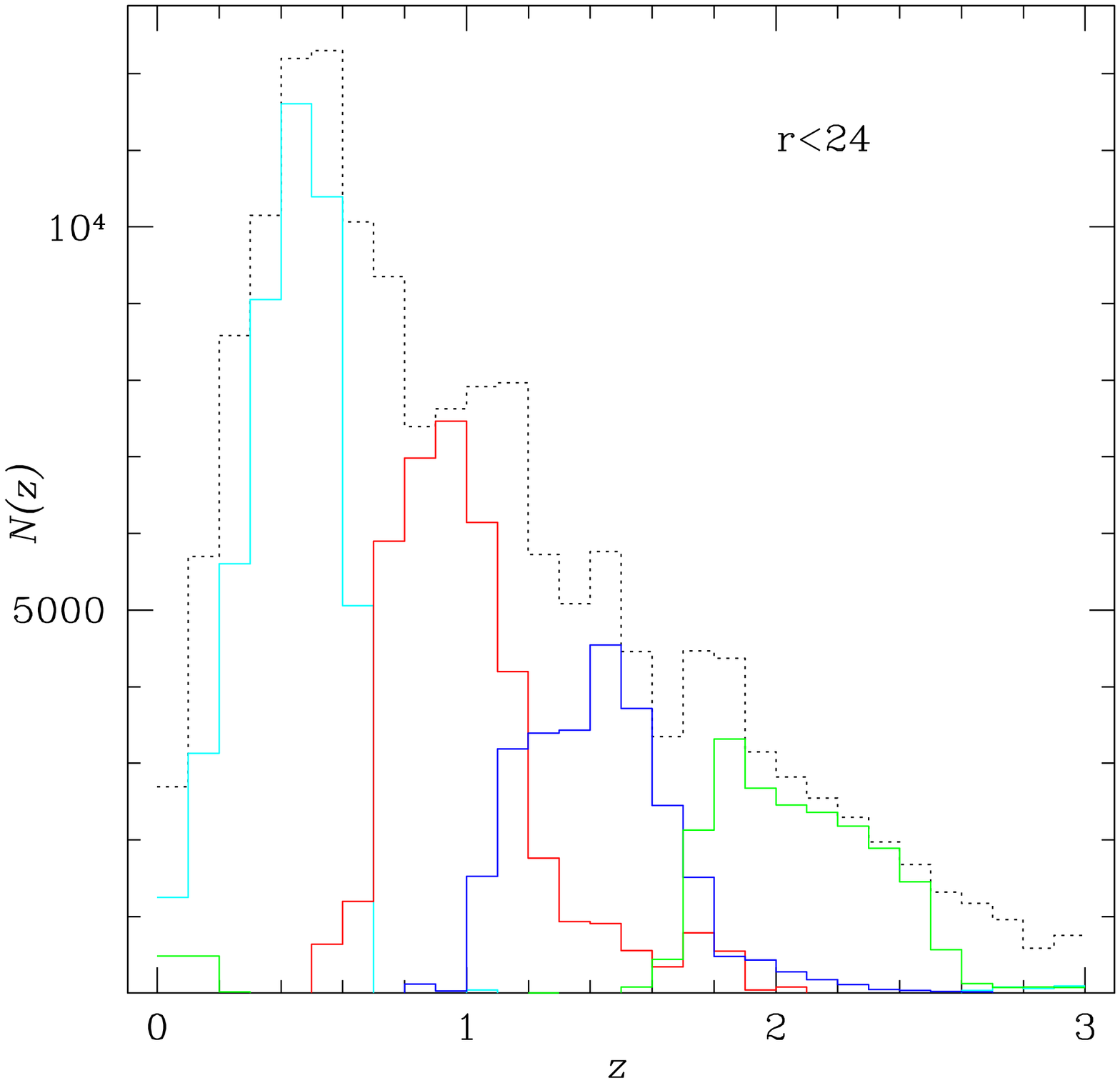}{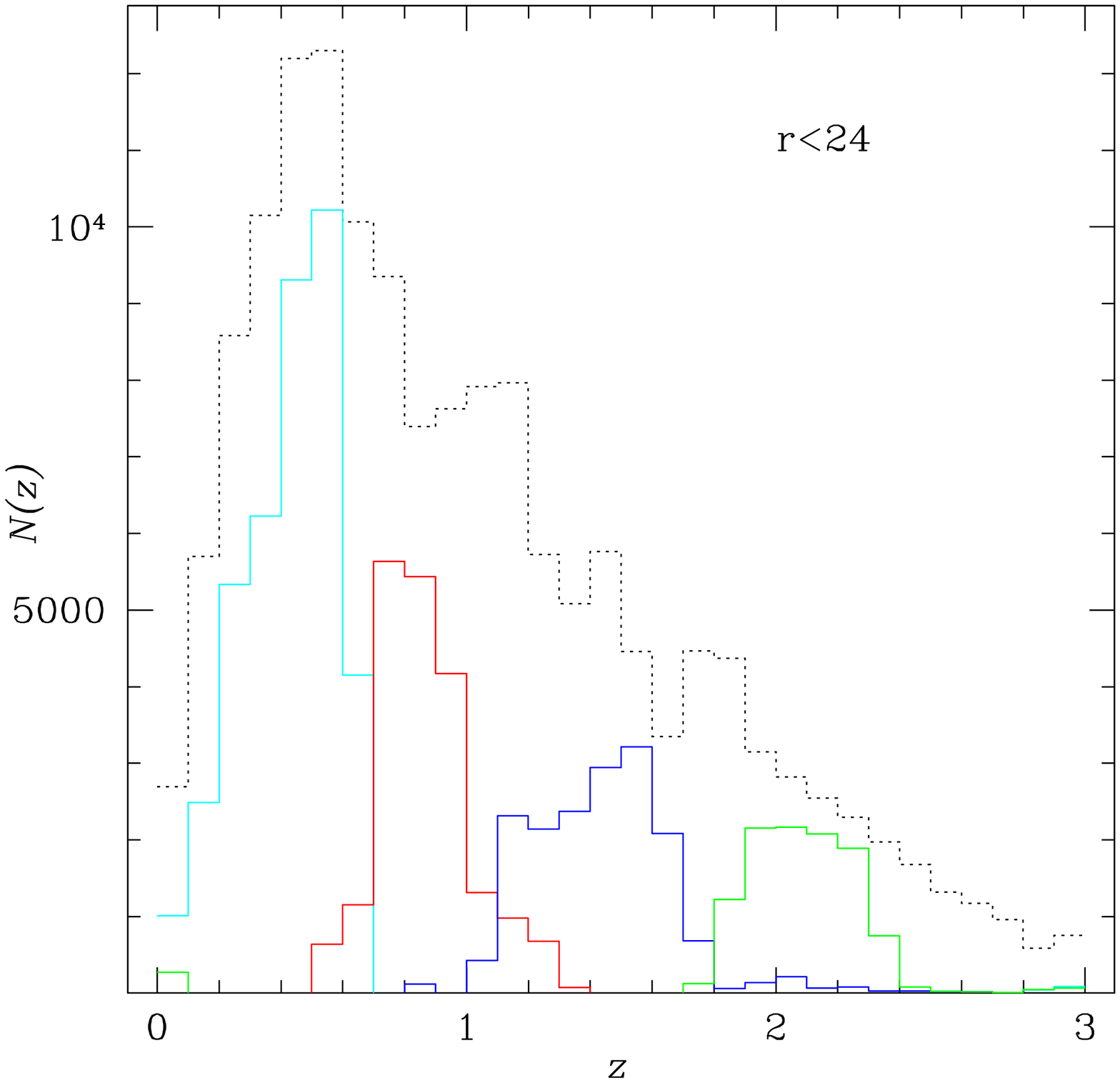}
\caption{\small As in Figure \ref{fig:dndz}, but for galaxies with 
$r<24$. 
}
\label{fig:dndz2}
\end{figure}

\begin{figure}[t]
\plottwo{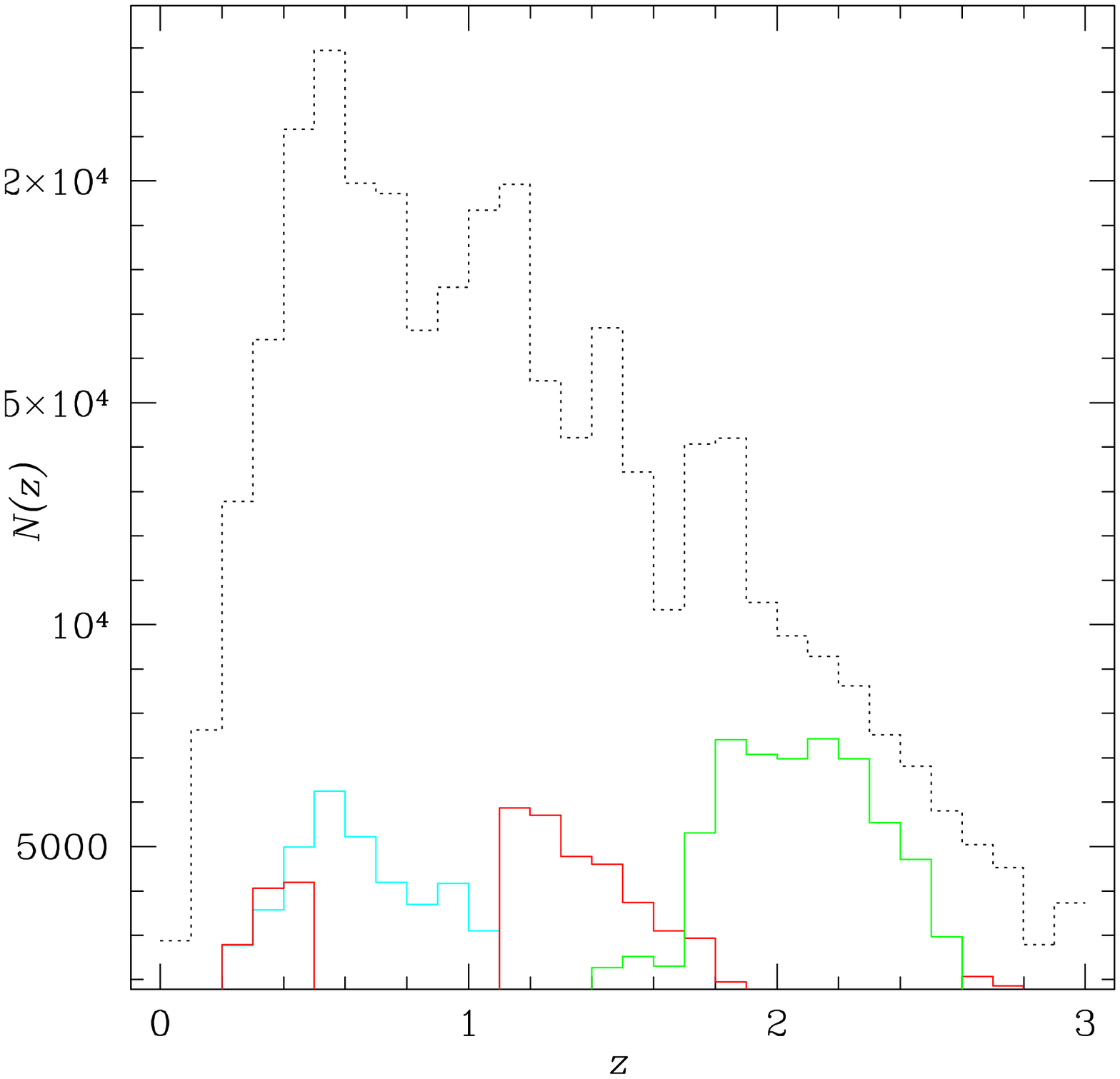}{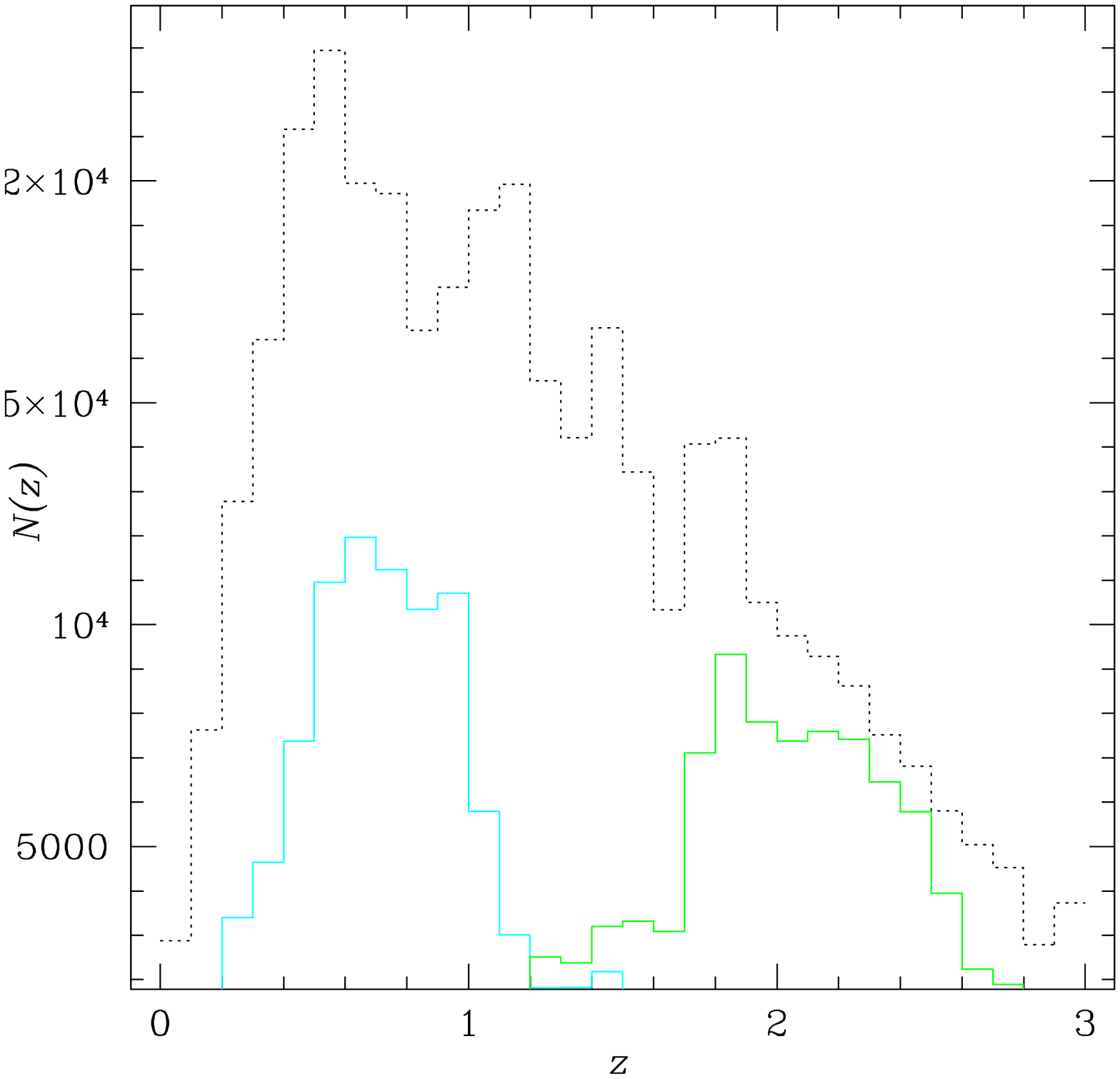}
\caption{\small As in Figure \ref{fig:dndz}, but using only $r-i$ color and
  $i$ magnitude to make the cuts. The left panel shows three
  sub-samples (with significant overlap) and the right panel two well
  separated sub-samples. 
}
\label{fig:dndz3}
\end{figure}

\section{Color tomography with two and three-band imaging}
\subsection{Redshift sub-samples with color cuts}

Having constructed our mock galaxy catalog, we used a set of heuristic
criteria to cluster galaxies in color space. The goal is to construct
2-4 sub-samples with the ``cleanest'' possible redshift
distributions. To achieve this we identify the redshift distribution of
each region of color space, discard the regions with poorly behaved
distributions, and group the rest into a set of sub-samples that would
be useful for lensing tomography, i.e. have well separated and compact
redshift distributions.

We pixelized $g-r$ vs. $r-i$ space into 400 pixels, and used the
redshift distribution of each pixel to rank pixels in order of
increasing mean redshift. We also characterized the compactness of the
distribution associated with each pixel by computing its low order
moments.  We then grouped pixels into sub-samples following a set of
heuristic criteria:
\begin{itemize}
\item The mean redshift associated with each pixel in a sub-sample 
fell within a well-defined redshift range. 
\item The redshift distribution of each pixel retained 
was sufficiently compact (had a small  variance). 
\item The joint distribution of pixels included a desired fraction 
of the total number of galaxies (typically more than a quarter of the total). 
\item The mean redshifts of the 3 or 4 sub-samples were useful for 
lensing tomography, i.e. were sufficiently well spread out over the range 
$0.3\lsim z\lsim 2.5$. 
\end{itemize}
This was iterated to arrive at the final selection and grouping of color 
pixels and thus the boundaries of the final redshift sub-samples. 

Figure \ref{fig:dndz} shows the results of making color cuts to isolate
four redshift intervals with boundaries at $z = 0, 0.7, 1.3, 2.0,
2.8$. We can trade-off number of galaxies within a sample versus how
``clean'' the redshift intervals are: the left and right panels of the
figure show the results of different trade-offs. In the right panel,
about half of the galaxies are discarded so that the resulting redshift
samples have almost no overlap. In the left panel, the overlap is
at the $10\%$ level or smaller. This level of overlap provides a clean
enough separation for WL tomography, but risks contamination of the
lensing signal by intrinsic alignments. Hence we use the more
conservative choice in the right panel for dark energy forecasts below.
We also note that for baryon oscillation measurements the widths in
redshift of these sub-samples are too large; one would need to focus on
finding particular galaxy types that yield tighter redshift
distributions.

Figure \ref{fig:dndz2} shows the results for a shallower sample with
limiting magnitude $r=24$. The basic result is similar, though with
fewer galaxies at $z>1$, it is more difficult to make four well
separated sub-samples.

We also explored the possibility of lensing tomography with just two-band
imaging. We used cuts in $r-i$ color and $i$ band magnitude to create
two or three different redshift distributions (note that for the three-band
case, we did not use the magnitudes to improve our redshift
selection). While this is clearly an idealized exercise, in that it
assumes a large calibration sample and does not include systematic
errors or allow for variation in the spectral templates, the results in
Figure \ref{fig:dndz3} show that it is not a hopeless goal. Tomography
is possible in up to three bins if $\sim 80\%$ of the galaxies can be
discarded (left panel), though with significant overlap between the
first two redshift samples. An alternative is to make just two samples
(right panel) while losing fewer galaxies and having minimal overlap
between the samples. We will quantify the consequences for
cosmological parameter estimation below. 

\subsection{Lensing power spectra}

\begin{figure}[t]
\epsscale{0.7}
\plotone{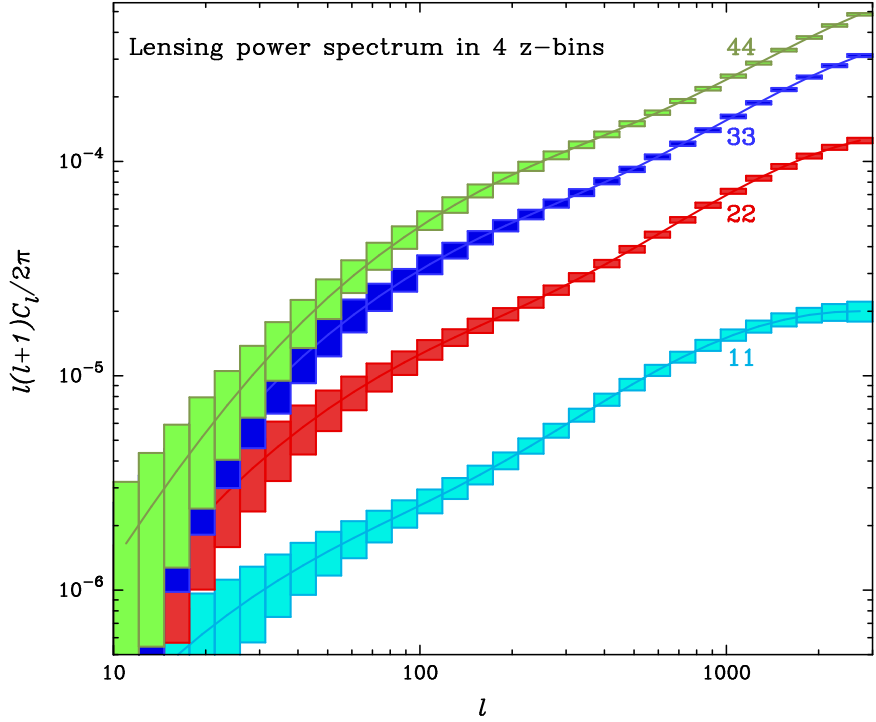}
\caption{\small Lensing power spectra $C(\ell)$ 
versus wavenumber $\ell$ from four redshift sub-samples, with
boundaries at $z=0, 0.7, 1.3, 2.0, 2.8$. The error bars include sample
variance and shot noise due to intrinsic ellipticity error for a
$2,000$ square degree survey with $40$ galaxies per square arcminute. 
Note that the Fisher errors on dark energy parameters also include all
the cross-spectra, which are not shown here. 
}
\label{fig:power}
\end{figure}

\begin{figure}[t]
\epsscale{1.1}  
\plottwo{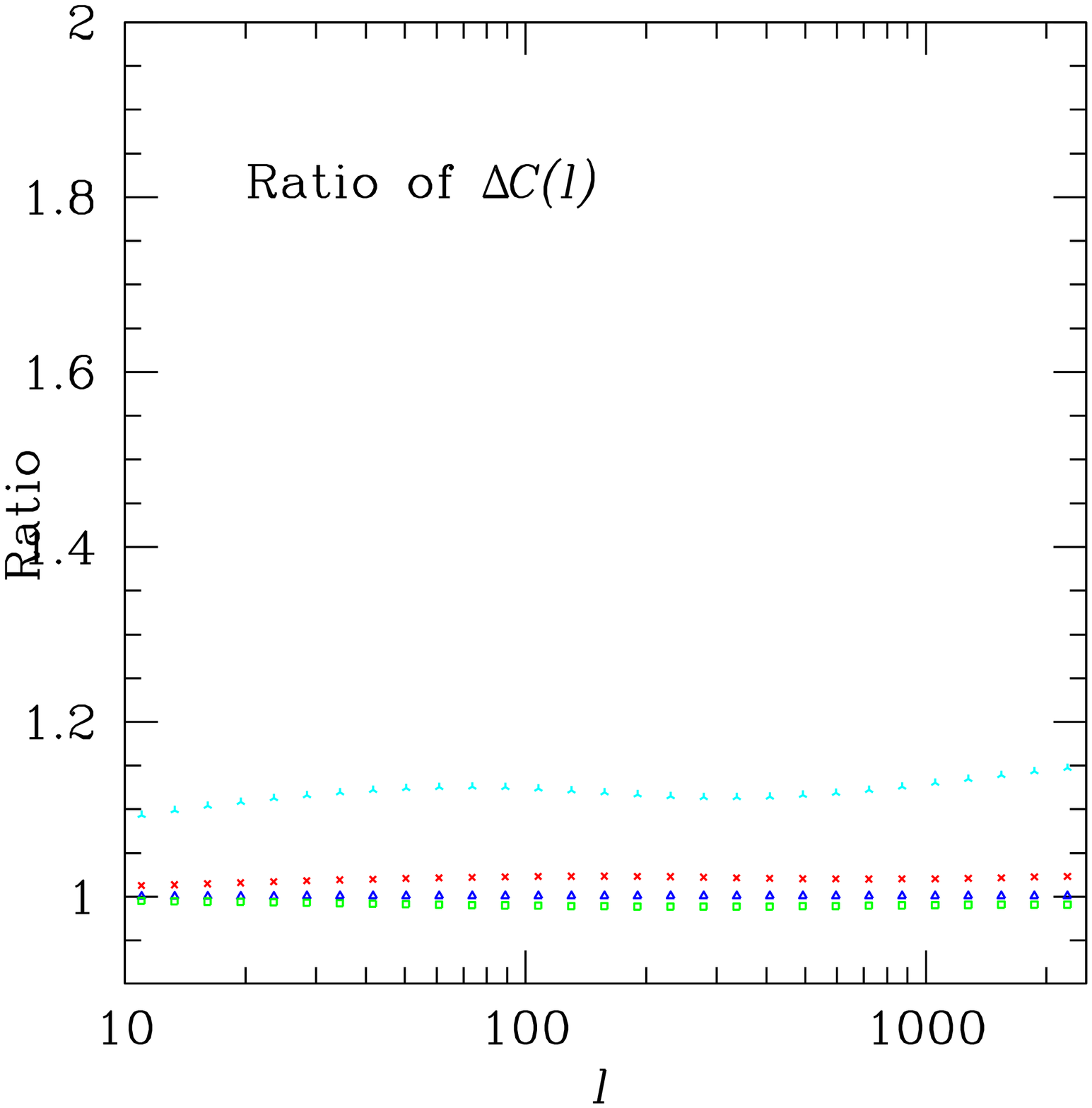}{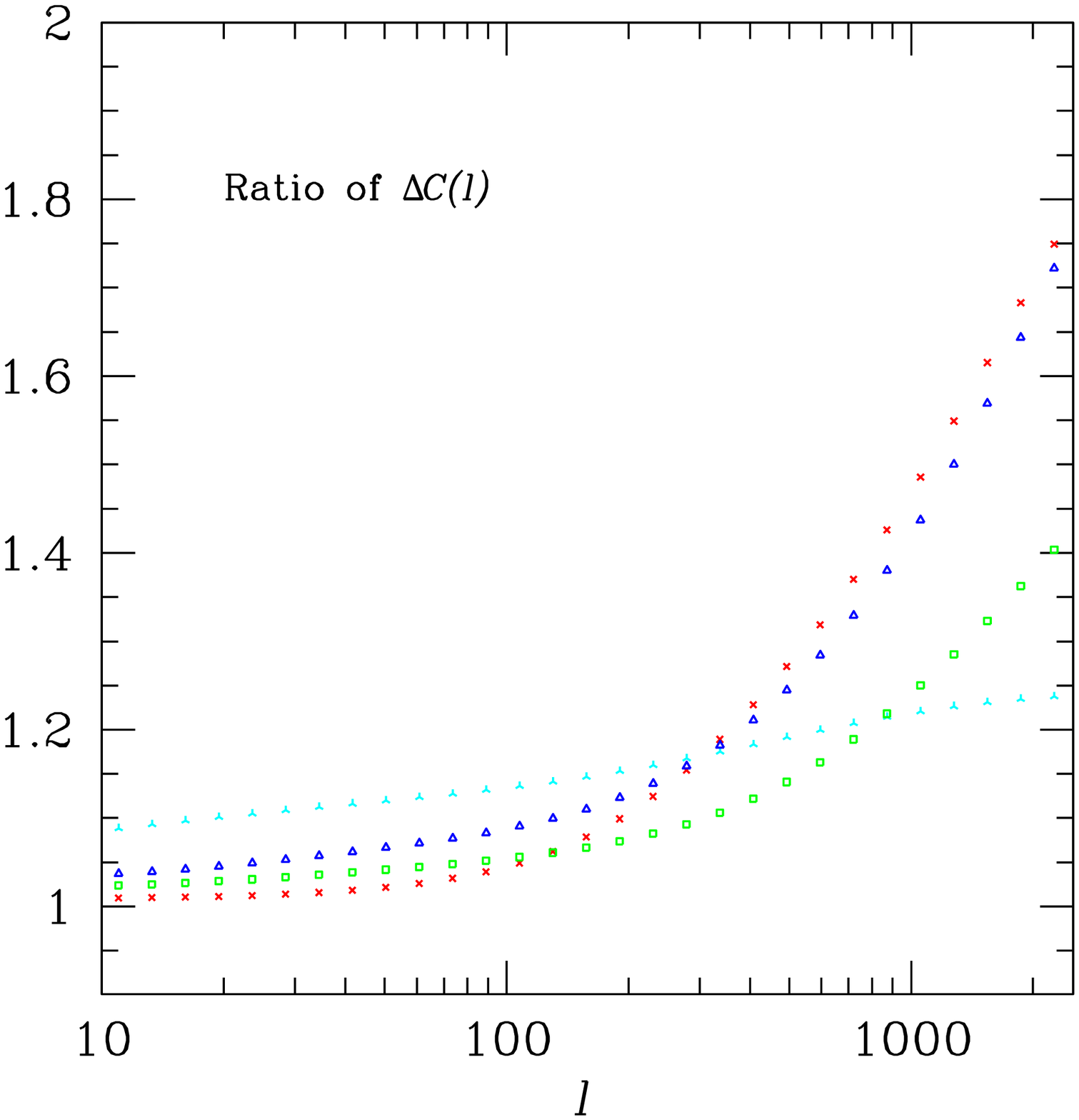}
\epsscale{}  
\caption{\small Ratio of power spectrum errors with idealized photo-z's 
and with color cuts. The power spectrum errors with photo-z's 
assume that all galaxies within
a given redshift range are used for the lensing power spectrum. 
The errors with color-cut are computed from the redshift distributions 
shown in Figure \ref{fig:dndz}; these are larger as some fraction of 
galaxies have been discarded. The two panels use the distributions 
in the left and right panels in Figure \ref{fig:dndz} respectively; 
the colors of the symbols above match those of the distributions.  
The ratio of errors increase with $\ell$ since the relative contribution of 
the shot noise term to the error is larger at high $\ell$. 
}
\label{fig:power_ratio}
\end{figure}

Figure \ref{fig:power} shows the lensing power spectra with errors for a
2,000 square degree survey with three-band imaging, using the four
redshift distributions of Figure \ref{fig:dndz}. Next we compare the errors on
the power spectra for two cases: (a) All galaxies within a redshift
range were used to compute the power spectrum (this represents the 
case of idealized photo-z's), and, (b) Only
galaxies selected using color cuts were used to compute the power 
spectrum (with distributions shown in the left panel of Figure
\ref{fig:dndz}). 
While the sample variance contribution is nearly the same as it
depends on the amplitude of the power spectrum itself, the shot noise
contribution is higher with the color cuts due to the loss in number 
density of galaxies. 

Since the change in errors is undetectable on the power spectrum plot,
we show the ratio of the errors for the two cases (idealized photo-z's 
vs. color cuts) in Figure \ref{fig:power_ratio}. The left and right 
panels correspond to the two sets of color cuts used in 
Figure \ref{fig:dndz}.  The 
errors increase by a few percent to over 50\% depending on the sample
and the range of $\ell$. At low wavenumber, $\ell\sim 100$, for all
cases the increase in error is at the 10\% level. Thus color
tomography does not degrade the errors at low wavenumber, since the errors
are dominated by sample variance. At high wavenumber $\ell\sim
1000$ the degradation depends on the fraction of galaxies discarded in
making the redshift sample; for the right panel it is as high as
50\%. The impact on dark energy parameters depends on the relative weights 
at low and high $\ell$. Note that we have underestimated the sample
variance contribution to the errors by using only the Gaussian
terms. The degradation due to the shot noise contribution 
would be smaller if non-Gaussian contributions to the sample variance
were included (these are significant at $\ell\gsim 1000$).

\subsection{Fisher analysis for cosmological parameters}

%%%AJC all of the techniques are photoz of some sort
\begin{table}
\begin{center}
Cosmological Parameters: Lensing\\ 
\begin{tabular}
{l\colskip ccc\colskip ccc\colskip ccc} \hline
&\multicolumn{3}{c}{\hspace{-8em}{\it Photo-z's}}
&\multicolumn{3}{c}{\hspace{-12.0em} {\it 3 bands/ 2 colors}}
&\multicolumn{3}{c}{\hspace{-4.5em}{\it 2 bands/1 color}}\\
& 6 $z$-bins & & 4 $z$-bins & 4 $z$-bins$^*$  & & 3 $z$-bins & 2 $z$-bins\\ 
\hline
$\sigma(\Omega_{\rm de})$
& 0.019& & 0.023 & 0.027&  &0.048  &0.1\\
$\sigma(w_0)$
&0.19& &   0.22&   0.25&  &0.46 &0.97&\\
$\sigma(w_a)$
& 0.64& &  0.72 &  0.86&  &1.4& 2.6&\\
%$\sigma(n_s)$
%& 0.20& &  0.12&   0.16& & 0.27& 0.23&\\
\hline
\end{tabular}
\end{center}
\caption{Summary of parameter constraints from lensing 
tomography using photo-z's with 6 redshift bins (column 2);
color cuts with three-band imaging
to make 4 $z$-bins (column 3); 
and 4 $z$-bins  with minimum overlap (denoted by $^*$, column 4); 
color cuts with two-band imaging to make 3 $z$-bins (column 5);   
and 2 $z$-bins with minimal overlap (column 6).   
All errors are $68\%$ confidence-level errors 
and include marginalization over the other  parameters. 
Note that we have used $f_{\rm sky}=0.05$ 
and all the errors scale as $\propto f_{\rm
 sky}^{-1/2}$.}  
\label{tab:lens}
\end{table}

\begin{table}
\begin{center}
Cosmological Parameters: Lensing+CMB\\ 
\begin{tabular}
{l\colskip ccc\colskip ccc\colskip ccc} \hline
&\multicolumn{3}{c}{\hspace{-8em}{\it Photo-z's}}
&\multicolumn{3}{c}{\hspace{-12.0em} {\it 3 bands/ 2 colors}}
&\multicolumn{3}{c}{\hspace{-4.5em}{\it 2 bands/1 color}}\\
& 6 $z$-bins & & 4 $z$-bins & 4 $z$-bins$^*$  & & 3 $z$-bins & 2 $z$-bins\\ 
\hline
$\sigma(\Omega_{\rm de})$
& 0.013& & 0.014& 0.016& & 0.025 & 0.033&\\
$\sigma(w_0)$
&0.11& & 0.13& 0.15& & 0.24 & 0.32&\\
$\sigma(w_a)$
&0.27& & 0.30& 0.36& & 0.54 & 0.73&\\
%$\sigma(n_s)$
%&0.0030& &0.0030& 0.0030& & 0.0030 & 0.0031&\\
\hline
\end{tabular}
\end{center}
\caption{Summary of parameter constraints from lensing 
 tomography with CMB Planck priors. The columns are as in Table
 \ref{tab:lens}. }
\label{tab:lens+cmb}
\end{table}

Having computed the lensing power spectra and the errors on them for 
different redshift samples, we estimate the errors on dark energy 
parameters using the Fisher matrix formalism.  
This formalism assesses how well given
observables can distinguish the true (``fiducial'') cosmological model
from other models.  The parameter forecasts we obtain depend on the
fiducial model and are also sensitive to the choice of free
parameters. We include all the key parameters that may affect lensing
observables within the CDM and dark energy cosmological framework: 
the density parameters are $\Omega_{\rm de}(=0.73)$,
$\Omega_{\rm m}h^2(=0.14)$, and $\Omega_{\rm b}h^2(=0.024)$ (note that
we assume a flat universe); the primordial power spectrum parameters are
the spectral tilt, $n_s(=1)$, the running index, $\alpha_s(=0)$, and the
normalization parameter of primordial curvature perturbation,
$\delta_\zeta(=5.07\times 10^{-5})$ (the values in the parentheses
denote the fiducial model). We employ the transfer function of matter
perturbations, $T(k)$, with baryon oscillations smoothed out (Eisenstein
\& Hu 1999). 
The dark energy equation of state parameters are
 $w(a)=w_0+w_a(1-a)$, with fiducial values $w_0=-1$ and $w_a=0$. 

Combining weak lensing with constraints from CMB
temperature and polarization anisotropies can be a powerful way to lift
parameter degeneracies (e.g. Hu \& Tegmark 1999; Takada \& Jain 2004).  
When computing
the Fisher matrix for the CMB, we employ 9 parameters: the 8
parameters above plus the Thomson
scattering optical depth to the last scattering surface, $\tau(=0.10)$.
The Fisher matrix for the
joint experiment is given by adding the CMB Fisher matrix to the lensing
Fisher matrix as $F_{\alpha\beta}=F^{\rm WL}_{\alpha\beta} +F^{\rm
CMB}_{\alpha\beta}$.  We  ignore the contribution to the CMB
from the primordial gravitational waves.  We use the publicly-available
CMBFAST code \citep{cmbfast} to compute the angular power spectra of
temperature anisotropy, $C^{\rm TT}_l$, $E$-mode polarization, $C^{\rm
EE}_l$, and their cross correlation, $C^{\rm TE}_l$.  Specifically we
consider the noise per pixel and the angular resolution of the Planck
experiment that were assumed in Eisenstein et al. (1998).  Note that we
use the CMB information in the range of multipole $10\le l\le 2000$, and
therefore we do not include the ISW effect at low multipoles $l\simlt
10$ which might be affected by dark energy perturbations.

\subsection{Tomography with idealized photo-z's versus color cuts}

We compare the Fisher errors for lensing tomography with color cuts
versus what is achievable with six redshift bins derived from an
idealized set of photo-z's using the LSST filter set. In both cases we
ignore the scatter of galaxies across bins due to photometric errors and
inexact spectral templates for galaxies. For color cuts, we consider
three and two band imaging, and in each case we use two sets of redshift
distributions to represent a more and less conservative treatment. These
correspond to the right (more conservative) and left panels of Figures
\ref{fig:dndz} (three band imaging) and \ref{fig:dndz3} (two band
imaging).  We consider constraints from lensing alone as well as from
lensing with CMB priors. We focus on dark energy parameters $w_0$ and
$w_a$ but also comment on the errors on the primordial power spectrum.

\begin{figure}[t]
\plotone{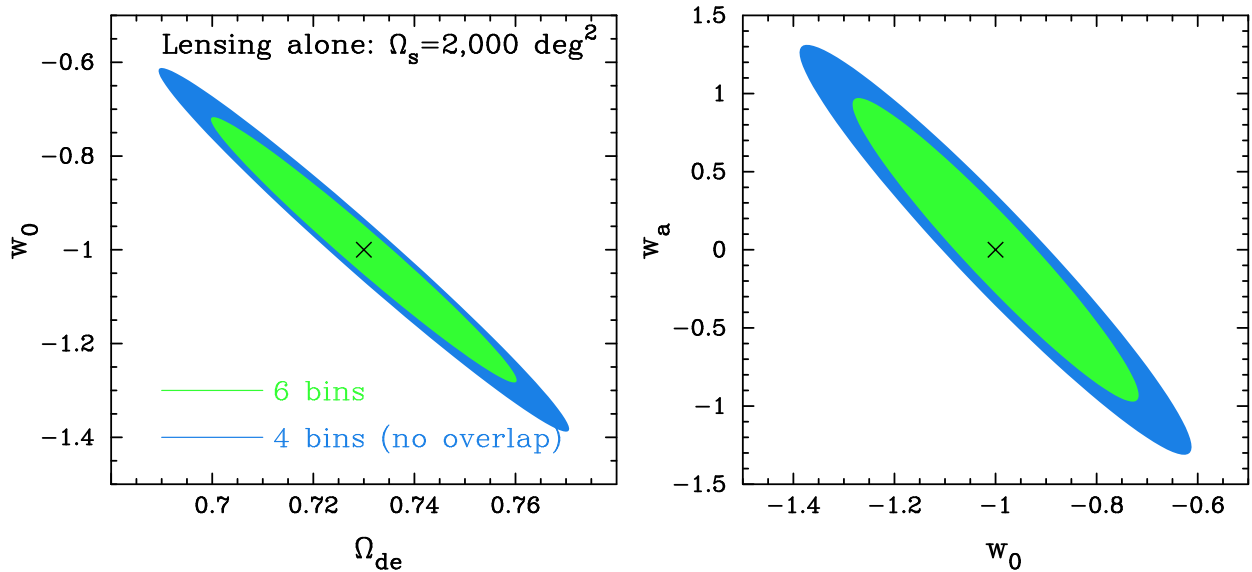}
\caption{\small Comparison of Fisher errors on dark energy parameters
using ``idealized photo-z'' tomography (using all galaxies and 6
redshift bins - green ellipse) versus color tomography (blue ellipse).
The color tomography ellipses are based on the four redshift sub-samples
made with $g-r$ and $r-i$ cuts shown in the right panel of Figure
\ref{fig:dndz} (the more conservative choice).  }
\label{fig:fisher1}
\end{figure}

\begin{figure}[t]
\plotone{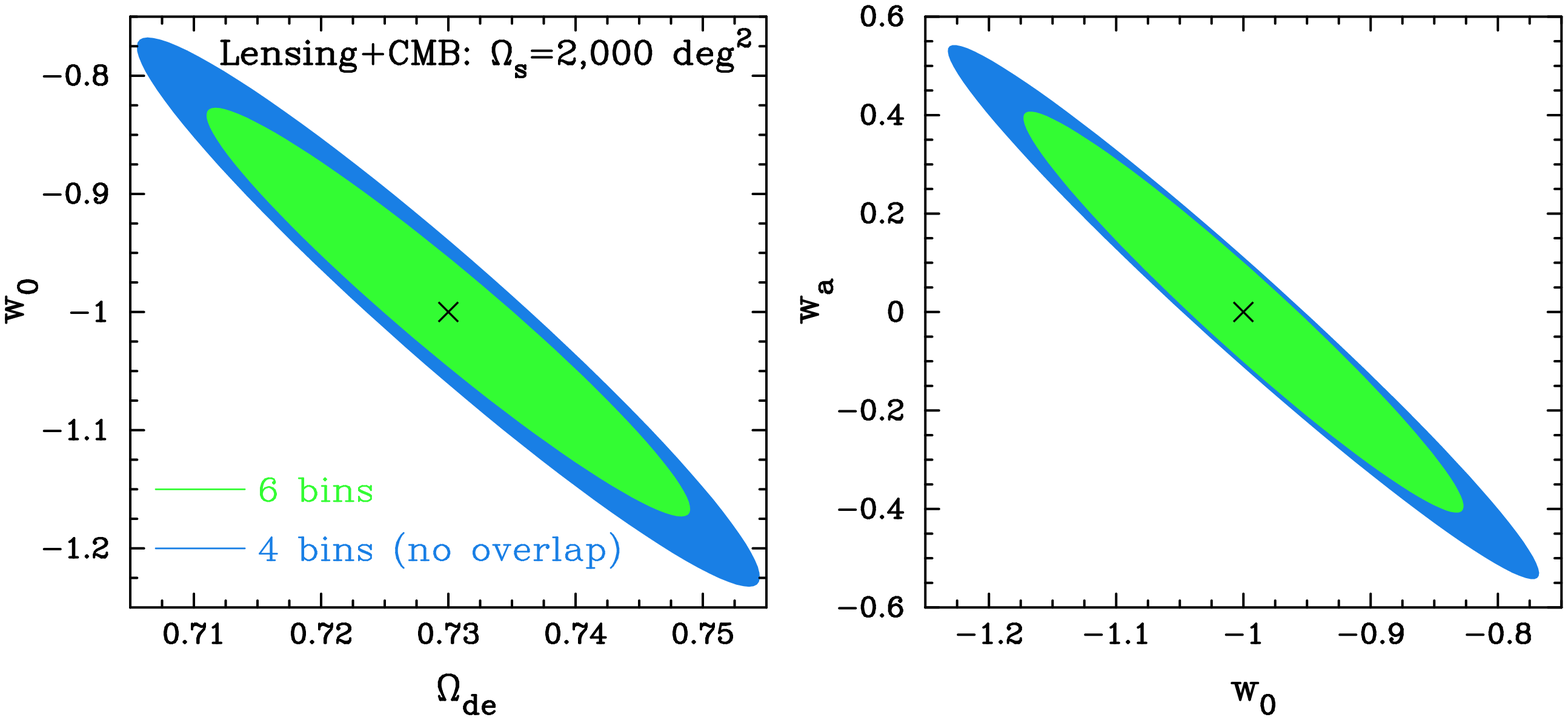}
\caption{\small As in Figure \ref{fig:fisher1}, but including Planck
  priors for the CMB. This reduces the degradation for color tomography.
  }
\label{fig:fisher2}
\end{figure}

Figures \ref{fig:fisher1} and \ref{fig:fisher2} show the results for the
dark energy parameters while Tables \ref{tab:lens} and
\ref{tab:lens+cmb} give the numerical values of the errors on
$\Omega_{\rm de}$, $w_0$, $w_a$.
These are marginalized over all other parameters. Table
\ref{tab:lens} shows the constraints from lensing, while Table
\ref{tab:lens+cmb} includes the Fisher errors expected from CMB data
from the Planck satellite. We have used only the lensing power spectrum;
adding information from the lensing bispectrum or lensing-galaxy
cross-correlations would improve the parameter constraints.
 
In Tables \ref{tab:lens} and \ref{tab:lens+cmb} the errors in column 2
(from 6 redshift bins with idealized photo-z's) can be compared with
columns 3 and 4 (4 $z$-bins using color cuts from three-band imaging data)
and columns 5 and 6 (3 and 2 $z$-bins from two-band imaging). Columns 4
and 6 represent the conservative option for three and two-band imaging
respectively, in that the redshift sub-samples have minimal overlap. For
three-band imaging, even the conservative choice of column 4 shows that
the degradation is modest: at the 30\% level compared to the ideal case
of 6-bin tomography. The loss of about half the galaxies does not prove
very damaging as much of the cosmological information comes from the
sample variance dominated regime. For two-band imaging, the degradation
depends on whether one uses 3 or 2 $z$-bins. The last column shows that
there is significant degradation with 2 $z$-bins. We found that even for
this case, one parameter representing $w$ at the pivot redshift can be
measured well. But constraints on $w_0$ and $w_a$ separately are
severely compromised with just 2 $z$-bins.

The parameter degradations are significantly reduced when CMB
information is used, as shown in Table \ref{tab:lens+cmb}. This is due
to the coarser requirements of lensing tomography once the high-$z$
information from the CMB is used. Note that all these results depend on
the lensing survey size, as the shot noise contribution affects
parameter errors more for a smaller survey. And the results in Table
\ref{tab:lens+cmb} are less impressive if the CMB priors are from WMAP
instead of Planck.

The constraints on the three dark energy parameters are shown in Figures
\ref{fig:fisher1} and \ref{fig:fisher2}. Idealized photo-z tomography is
compared with 4 $z$-bin tomography from three-band
imaging, using the more conservative choice of redshift sub-samples. 
These confirm the information in the tables: the
degradation in the dark energy parameters is modest, and is even
smaller in the best determined combination of $w_0$ and $w_a$, which 
is the value of $w$ at the pivot redshift for lensing. 

We also note that lensing primarily probes parameters that are more
sensitive to the lensing power spectrum amplitude rather than its shape
parameters (see the discussion in Takada \& Jain 2004). For our case
these parameters are $\Omega_{\rm de}$, $w_0$, $w_a$ and $\delta_\zeta$.
Adding redshift slices for tomography improves the errors on these
parameters significantly up to 4-6 slices 
as it provides information on the redshift evolution of the
amplitude of the lensing power spectrum. 
Errors on the shape parameters, such as the logarithmic slope of the
power spectrum $n_s$, do not change appreciably. 
%degrade because of the increased
%shot noise in the narrower redshift slices. As the CMB provides an
%accurate measurement of $n_s$, adding lensing information does not
%improve it. 
%This leads to the counter-intuitive result that, with
%lensing alone, the accuracy on $n_s$ improves with coarser $z$-bins (see
%Table \ref{tab:lens}). This is a result of the shot noise contribution
%affecting the accuracy on the shape of the power spectrum less, while
%the evolution of the amplitude is affected more, leading to poorer dark
%energy constraints.

\section{Discussion}

We have studied the prospects for lensing tomography with a wide area
imaging survey that has limited color information. While it is never
desirable to limit color information, it may be that a full complement
of filters is not available at a survey telescope, that it is
un-realistic to image the full survey area in all filters, or that
additional science requirements for a survey dictate an observing
strategy that is not optimal for lensing tomography. The questions we
address in this study are whether lensing tomography is at all feasible
with two or three band imaging data, and if so, what is the trade-off
between filter choice and sky coverage.

We use the fact that lensing tomography makes only statistical use of
estimated redshifts and, given the width of the lensing kernel, the
redshift bins can be quite broad (though they must be known
accurately). Further, for planned wide-area surveys  
lensing measurements are not shot noise limited -- hence a
significant fraction of galaxies can be discarded if their redshifts are
ill-defined without severely degrading parameter accuracy. With these
considerations in mind, we generate a mock catalog of galaxies extending
to high redshift with known types, redshifts and colors.  This provides
an estimate of the redshift distributions for each part of $g-r$ and
$r-i$ color space. From this we select regions in color space that
produce sub-samples with redshift distributions well suited for
tomography.  Errors on the power spectra in different redshift bins 
from this color tomography are
then compared with what would be expected with idealized photo-z's.

The resulting degradation in the accuracy of cosmological parameters 
is shown in Tables \ref{tab:lens} and \ref{tab:lens+cmb}, and Figures
\ref{fig:fisher1} and \ref{fig:fisher2} for dark energy parameters
from color tomography. 
With three-band imaging, the errors on dark energy parameters 
are only modestly degraded compared to
idealized six-band photo-z's. Even with two-band imaging, lensing
tomography may be feasible for the case of high signal-to-noise
data and an adequate calibration sample. This result may have 
implications for how to trade-off filter choice and sky coverage 
for large area surveys and how to optimize survey strategy to maximize the 
scientific returns in its initial stages. For example, 
imaging in three filters instead of six
is more than twice as efficient in terms of survey time since the $u$
and $y$ filters (on the blue and red end) are substantially less
efficient. Thus, for a fixed survey duration, the survey area for a six
band program would be less than half that of the three band survey, with
all parameter errors scaling as $f_{\rm sky}^{-1/2}$. Such trade-off
studies must be carried out for the specific parameters and available
choices in observing strategy for a given survey; 
the fiducial parameters we have considered do not correspond 
to any real or planned survey known to us. 

For color tomography an adequate calibration sample is essential. The
calibration will likely require a two-step approach: a sample of
spectroscopic redshifts of over $\sim 10^4$ galaxies (for a $\sim 1000$
square degree survey), and a larger
sample of multi-band imaging with a full set of optical and possibly
infra-red filters.  Imaging to a depth equivalent to $r \sim 25$ in
say six bands over $10$-$20$ square degrees would provide a sample of well 
measured photo-z's for over $1$ million galaxies. 
This would enable us to map the redshift distributions over the desired
color space: e.g., it could provide photo-z's of 
over $10^4$ galaxies in each bin of $\sim 0.1\times 0.1$ magnitude 
in two-colors (assumed to be available for
the full survey area).  It has been recognized that even with five or 
six band imaging data, photo-z's would need to be carefully calibrated with
spectroscopic redshifts so that the means
and widths of redshift bins are known to high accuracy (Bernstein \&
Jain 2004; Huterer \etal \ 2006; Ma, Hu and Huterer 2006), and proposals
for calibration are being developed (Newman 2006). For color tomography
the calibration sample is even more important, as it forms the basis for
grouping and discarding regions in color space. We leave for future work
the detailed requirements and strategy for obtaining the calibration sample.

Our current study is based on estimates of statistical errors in the
lensing power spectra -- in future work we will include systematic
photometric uncertainties and other systematic errors that affect
lensing. These will include a realistic modeling of
photometric errors, filter design, and will address whether the spectral
templates are adequate for describing high-redshift galaxies. The
inclusion of these realistic errors is essential before one can make
detailed trade-off studies of survey duration and filter choice.
Several caveats are in order until such a study is done: systematic
errors may lower the signal-to-noise and this may depend on scale,
testing for systematics is more difficult with coarser redshift bins, 
and marginalizing over intrinsic ellipticity contributions may be harder
(see below). Finally, the inclusion of other statistics such as higher
order correlations and cross-correlations with the galaxy distribution
can alter the results on parameter accuracies.
Some of these issues can be tested on ongoing multicolor
surveys, such as the CFHT Legacy Survey. 

Our results on the power spectrum errors in Figure
\ref{fig:power_ratio} show that the degradation in
errors is worse at high $\ell$, because the shot noise term dominates on
small scales. This raises the question: should one choose different
galaxy samples at different $\ell$: be more conservative at low $\ell$,
since number density matters less, to minimize redshift overlap and
bias? This merits a detailed study as it has the potential to impact 
different survey strategies.  Such a study must include non-Gaussian 
contributions that increase the sample variance at high $\ell$; this 
in fact lowers the degradation for color tomography as the shot noise
regime shifts to higher $\ell$ (e.g. Kilbinger \& Schneider 2005). 

Finally, intrinsic ellipticity
alignments (e.g. Heymans \etal \ 2006) and ellipticity-shear alignments
(Hirata \& Seljak 2004) must be considered. While we did choose as our
fiducial sub-samples ones with well separated redshift
distributions, it may be necessary to only use
cross-spectra to eliminate intrinsic ellipticity contributions
(e.g. Takada \& White 2004).  This would increase the parameter errors
from having a smaller number of redshift bins. Similarly ellipticity-shear
correlations may need to be measured from the data and marginalized
over, which is easier to do with finer redshift binning. Four redshift
bins are the minimum needed to separately fit for both kinds of intrinsic
ellipticity correlations, this may impose a minimal requirement on the
needed color information. The impact of intrinsic alignments is
likely to be the most important issue for future studies related to
color tomography. 

\acknowledgments 

We thank Mike Jarvis for stimulating discussions at all stages of this
work and Eric Linder for valuable comments on an early draft. We thank
Gary Bernstein, Peter Schneider, Ravi Sheth and Fritz Stabenau for
helpful discussions. This work was supported in part by the COE program
at Tohoku University.  MT acknowledges support from a Grand-in-Aid for
Scientific Research (17740129) of the Ministry of Education, Culture,
Sports, Science and Technology in Japan. BJ is supported in part by NASA
grant NAG5-10924 and by the Research Corporation. 
AJC is supported in part by an NSF ITR award and NSF
CAREER award AST9984924. AJC would like to thank Google for their
hospitality while completing this work. We acknowledge use of the
publicly-available CMBFAST code.

\onecolumn

\end{document}